\documentstyle[preprint,aps,prb,tighten]{revtex}

\newcommand{\ea}{{\it et al}}

\newcommand{\ee}{\mbox{{\boldmath $\varepsilon$}}}
\newcommand{\kk}{\mbox{{\boldmath $k$}}}
\newcommand{\rr}{\mbox{{\boldmath $r$}}}
\newcommand{\qq}{\mbox{{\boldmath $q$}}}
\newcommand{\RR}{\mbox{{\boldmath $R$}}}
\newcommand{\RRang}{\mbox{{\boldmath $\hat R$}}}
\newcommand{\Prr}{\mbox{$P(\mbox{\boldmath $r$})$}}

\newcommand{\ez}{\mbox{{\boldmath $\varepsilon$}$\parallel \!\! {z}$}}
\newcommand{\exy}{\mbox{{\boldmath $\varepsilon$}$\parallel \!\! xy$}}
\newcommand{\eaa}{\mbox{{\boldmath $\varepsilon$}$\parallel \!\! {a}$}}
\newcommand{\ecc}{\mbox{{\boldmath $\varepsilon$}$\parallel \!\! {c}$}}

\newcommand{\qqang}{\mbox{{\boldmath $\hat q$}}}
\newcommand{\nnang}{\mbox{{\boldmath $\hat n$}}}
\newcommand{\kkang}{\mbox{{\boldmath $\hat k$}}}
\newcommand{\rrang}{\mbox{{\boldmath $\hat r$}}}
\newcommand{\eeang}{\mbox{{\boldmath $\hat \varepsilon$}}}
\newcommand{\rn}{\mbox{$R_{N}^{(j)}$}}

\newcommand{\psik}{\mbox{$\psi^{(-)}_{\bbox{k}}({\mbox{\boldmath $r$}})$}}

\newcommand{\estd}{\text{e}}
\newcommand{\istd}{\text{i}}
\newcommand{\dstd}{\text{d}}

\newcommand{\kedge}{{\it K} edge}
\newcommand{\tis}{\mbox{TiS$_{2}$}}
\newcommand{\tio}{\mbox{TiO$_{2}$}}

\newcommand{\oeq}{\mbox{O$_{\text{eq}}$}}
\newcommand{\oax}{\mbox{O$_{\text{ax}}$}}
\newcommand{\tic}{\mbox{Ti$_{\pm c}$}}

\newcommand{\pj}{\mbox{$P^{(j)}$}}
\newcommand{\dpjat}{\mbox{$\Delta P^{(j)}_{\text{ato}}$}}
\newcommand{\dpjdos}{\mbox{$\Delta P^{(j)}_{\scriptscriptstyle \text{DOS}}$}}

\newcommand{\xra}{x-ray absorption}
\newcommand{\phd}{photoelectron diffraction}
\newcommand{\pd}{probability density}

\newcommand{\prevb}[1]{Phys. Rev. B {\bf #1}}
\newcommand{\prevl}[1]{Phys. Rev. Lett. {\bf #1}}

\newcommand{\jpcm}[1]{J. Phys.: Condens. Matter {\bf #1}}

\newcommand{\mm}[1]{\mbox{$#1$}}
\newcommand{\bbt}[1]{\bibitem{#1}}

\RequirePackage[dvips]{graphics}

\begin{document}

\draft

\title{Spatial distribution of photoelectrons\\
 participating in formation of x-ray absorption spectra}

\author{ O.\ \v{S}ipr}
\address{Institute of Physics, Academy of Sciences of the Czech
Republic, Cukrovarnick\'{a}~10, 162~53~Praha~6, Czech~Republic}


\maketitle

\begin{abstract}
Interpretation of  x-ray absorption near-edge structure (XANES)
experiments is often done via analyzing the role of
particular atoms in the formation of specific peaks in the
calculated spectrum.  Typically, this is achieved by calculating the
spectrum for a series of 
trial structures where various atoms are moved and/or removed. A more
quantitative approach is presented here, based on comparing the
probabilities that a XANES photoelectron of a given energy can be
found near particular atoms. Such a photoelectron 
probability density can be consistently defined as a sum over squares
of wave functions which describe participating photoelectron diffraction
 processes, weighted by their normalized cross sections. A fine
structure in the energy dependence of these 
probabilities can be extracted and compared to XANES spectrum.
  As an illustration of this novel technique, we analyze the
photoelectron probability density at the Ti {\it K} pre-edge
of \tis\ and at the Ti \kedge\ of rutile \tio.

\vspace{2cm}

\begin{quote}
{\em To be published in Physical Review B, tentatively scheduled for 15 May
2002 issue.}
\end{quote}

\vspace{2cm}

\end{abstract}



\narrowtext

\section{Introduction}   \label{intro}
 
X-ray absorption fine structure (XAFS) is being used for studying both the
electronic and the real structure of solids.
For high photoelectron energies (\mm{E \gtrsim 200}~eV),
the extended x-ray absorption fine structure (EXAFS) can be
intuitively described in terms of backscattering of the excited
photoelectron by neighboring atoms and the EXAFS analysis has
become a standard tool for real structure investigations.  On the other
hand, x-ray absorption near edge structure (XANES) is lagging behind
in its applications for structural studies, despite some promising
applications in selected systems.\cite{pendry,cinani,bugaev,benfatto}
The reason for this rests in a more 
complex physics hidden behind XANES, resulting both in a more
difficult theoretical treatment (multiple-scattering, selfconsistency
in potentials, non muffin-tin effects) and in a lack of a proper
intuitive insight into the formation of XANES peaks.
The need for involving XANES in structural analysis stems partly from
the fact that EXAFS is predominantly sensitive only to atomic
distances and not to bond angles, and partly from the high
signal-to-noise ratio for some interesting classes of systems which
severely limits the ability to extract EXAFS oscillations from their
spectra. 

A means to interpret XANES in intuitively plausible terms
would, among others, facilitate application of XANES
spectroscopy in investigations of both the real and the electronic
structure.  
Various procedures were applied in the past with the aim to connect
XANES spectral features with real structure.  Among them, let us
mention inspecting the
effect of adding or removing certain atoms in the test
cluster,\cite{v2o5}  investigating the dependence of the
height of the pre-peak on the geometry of the nearest
neighborhood,\cite{farges}  calculating XANES 
 by summing over many scattering paths\cite{feff}
 or employing ``direct inversion'' technique for obtaining  
 atomic positions and scattering potential from experimental
XANES.\cite{migal} 
In this study, we would like to tackle the problem of interpreting
XANES spectra from yet another
side, namely, we want to explore the probability 
density of the photoelectrons.  Such a procedure can be ---
from a certain viewpoint --- considered as an answer to the naive
question ``where the XANES photoelectron really is''. 
Answering such a fundamental question would have implications for
real structure as well as for electronic structure studies.

At the beginning of this paper, a general consideration of the problem
will be outlined, the main goal being to define the task in exact
terms.  Then we will present few mathematical formula which describe
the problem and determine how the proper probability density ought to
be evaluated.  Next follows a discussion of some practical aspects of
evaluating the photoelectron probability density and the technique is
illustrated on the pre-edge structure of Ti {\it K} edge of \tis\ and
on the whole Ti {\it K} edge XANES of \tio.  Some more technical
details are given in the appendices.

\section{Localization of the photoelectron}

XAFS arises due to the energy-dependence of the core-electron
photoeffect: An electron absorbs an x-ray photon, is ejected off the
atom and starts to travel inside the solid.  In a stationary picture,
the wave function of the excited photoelectron can be viewed as being
subject to multiple scattering by neighboring atoms.  The probability
of photoabsorption oscillates with energy --- one can,
especially in the EXAFS regime, interpret this oscillatory behavior
intuitively as a consequence of either constructive or destructive
interference of the photoelectron wave function.  One can thus
associate a particular spectral peak with scattering of a
photoelectron of a certain energy.  The topic of this section (and of
this paper as a whole) is to explore the spatial localization of this
photoelectron.

\subsection{Wave function problem}
\label{sec_wave}

In the framework of the quantum theory, the popular question ``where
is the electron'' cannot be answered. However, it is possible to ask
what is the probability density $P(\rr)$\ that a 
quantum-mechanical object 
 can be found at a given place \rr\ ---  it is just
the square of the modulus of its wave function,
\begin{equation}
P(\rr) \; = \; \bigl| \psi(\rr) \bigr|^{2} \; \; \; .   \label{born}
\end{equation} 
So the problem is reduced  to the task of finding the wave
function $\psi_{\rm phe}(\rr)$\ which would appropriately describe the
photoelectron participating in the formation of XANES at
 a given energy. 

 The correct form of this wave function ought to
emerge from the way of calculating the \xra\ spectrum.
The x-ray absorption spectrum (XAS) intensity is proportional to the
probability $w$\ that a photon is absorbed by a core electron.  That 
can be expressed, within the first-order perturbation theory, as 
\begin{equation}
w \; = \; 2\pi \; \int \! \dstd\nu \: \bigl|\langle \psi_{\nu}|H_{I}|
\phi_{c} \rangle  \bigr|^{2} \; \delta(E_{c} + \omega - E_{\nu}) 
\; \; , \label{nusum}
\end{equation}
where $H_{I}$\ is the interaction hamiltonian perturbing the initial
electron state $|\phi_{c} \rangle$\  and the sum/integration
over $\nu$\ spans {\em any} complete set of electron
wave functions $|\psi_{\nu} \rangle$\ (Rydberg atomic units are used 
throughout this paper, taking \mm{m=1/2}, \mm{\hbar=1}, \mm{e^{2}=2},
\mm{c=2/\alpha}, \mm{\alpha=1/137.036}).  Applying standard procedures,
Eq.\ (\ref{nusum}) can be transformed into expressions 
which involve either sums over Bloch states,\cite{muller} or 
molecular cluster basis 
functions,\cite{dill} or  photoelectron diffraction
states\cite{lee} or which, in the case of 
Green function formalism, dispose of the final state wave functions
 altogether.\cite{vvedensky}  The important factor is that the choice
of the set of wave functions $|\psi_{\nu}\rangle$\
does not affect the outcome of Eq.\ (\ref{nusum}).  This is favorable
on the one hand, as 
one does not have to care about the particular form of wave functions 
 $|\psi_{\nu} \rangle$\ when calculating XAS intensity.  On the other
hand, it means that  the proper photoelectron wave function cannot 
just be borrowed from Eq.\ (\ref{nusum}) or from any of its clones. 

A close look at Eq.\ (\ref{nusum}) reveals that the situation is even
worse at the first sight.  Namely, it follows from Eq.\ (\ref{nusum})
that the total absorption rate is actually 
resultant from many {\em incoherent} processes.  Hence, there is simply
{\em no single wave function} 
like $|\psi_{\rm phe}\rangle$\ which could have been inserted into
Eq.\ (\ref{born}).  Nevertheless, one still can ask what is the
probability that any electron ejected as a results of an absorption of
a photon with energy $\omega$\ can be found at \rr.  One
only has to reformulate the problem slightly:  Instead of searching
for \mm{| \psi_{\rm phe}(\rr) |^{2}}, the quantity of interest should
rather be a weighted sum of 
probability densities of those wave functions which describe states
participating in the absorption process,\cite{stary} 
\begin{equation}
P(\rr) \; = \; \sum_{f} \, w_{f} \, \bigl| \psi_{f}(\rr) \bigr|^{2} \; \; .
\label{pr}
\end{equation}
The weights $w_{f}$\ with which the participating wave
functions contribute to \Prr\ are the probabilities that a core electron
$|\phi_{c}\rangle$\ is ejected into the state $|\psi_{f}\rangle$, 
\begin{equation}
w_{f} \; \sim \; 
\bigl| \langle \phi_{c}|H_{I}| \psi_{f}\rangle \bigr|^{2} \; \; .
\label{wf}
\end{equation}
They ought to be normalized so that their sum yields the total XAS
probability $w$.  Note that all this is
just another way of saying that the ejected photoelectron is described
not by a single quantum state but rather by a density matrix,
\begin{equation}
\label{denmat}
V = \sum_{f} w_{f} \: | \psi_{f}\rangle \langle \psi_{f}|
\; \; .
\end{equation}

Unfortunately, there appears to be a complication resulting from the
use of Eqs.\ (\ref{pr})--(\ref{wf}). 
Unlike in the case of Eq.\ (\ref{nusum}), the outcome of
Eqs.\ (\ref{pr})--(\ref{wf}) now, namely, {\em does depend} on the choice 
of the set of wave functions $|\psi_{f}\rangle$\ [complete on the energy
surface determined by the $\delta$-function in Eq.\ (\ref{nusum})].  There
is no way to fix this choice by relying solely on Eq.\ (\ref{nusum}).
It is a matter of physical intuition, not formal mathematical 
procedures, to establish which set of states $|\psi_{f}\rangle$\ is
relevant to the physical process in question.  A guide for this
choice could be the conservation of electron number:  As one 
has one electron in the initial state (it is the core electron
$|\phi_{c} \rangle$), one has to end up with one electron in the final
state as well.

The choice of states $|\psi_{f}\rangle$\ is quite straightforward in
the case of transitions to bound states (say, of a molecule).
The final states $|\psi_{f}\rangle$\ are just asymptotically decaying
bound states $|\psi_{b}\rangle$, normalizable so that 
\mm{\int \dstd^{3} \rr | \psi_{b}(\rr) |^{2} = 1}. 
The situation is more complicated for transitions to the
continuous part of the spectrum, as there is no {\em a priori}
preference for normalization and/or boundary conditions which such a
wave function ought to observe.

In order to find the correct wave functions, 
let us contemplate a finite cluster of atoms (a situation tacitly
assumed in most applications of the real-space
formalism)\cite{dill,lee,vvedensky} and follow 
the fate of the initially core electron. As a result of
absorbing a photon, this electron is
torn off an atom and, having its energy above the continuum threshold,
must 
finally leave the cluster and turn into a plane wave with a
well-defined momentum direction {\kkang}.  This means that  
the elementary incoherent processes involved in the sum (\ref{pr})
must be
{\em photoelectron diffraction} events.  Indeed, x-ray absorption is
conceptually nothing else but angularly integrated photoelectron
diffraction (PED).\cite{lee,fadley}  The final 
states $|\psi_{f}\rangle$\ are, therefore, time-inversed scattering
states, 
$|\psi_{\bbox{k}}^{(-)}\rangle$, which are in turn solutions of the
Lippman-Schwinger 
equation\cite{lee,bethe,nbd}
\begin{equation}
\psi_{\bbox{k}}^{(-)}(\rr) \; = \; 
    \estd^{\istd \bbox{k r}}
    \; + \; 
    \int \! \dstd^{3} \rr' \, 
    G^{(-)}_{0}(\rr,\rr') \, 
    V(\rr') \, 
    \psi_{\bbox{k}}^{(-)}(\rr') \; \; ,
\label{lse}
\end{equation}
where $G^{(-)}_{0}(\rr,\rr')$\ is the advanced free electron Green function
and $V(\rr')$\ describes the potential of the cluster.  The states
$|\psi_{\bbox{k}}^{(-)}\rangle$\ are normalizable to the delta function
\mm{\delta(\kk)} [apart from the constant factor \mm{1/(2\pi)^{3/2}}],
which guarantees that they 
describe exactly one electron at a time.\cite{bethe}

The probability density \mm{P(\rr)}\ obtained via 
Eqs.\ (\ref{pr})--(\ref{wf}) cannot be normalized to one.  However,
due to Eq.\ (\ref{lse}), 
it can be related to the probability density of a free electron, which
is described by the wave function \mm{\exp(\istd \kk \rr)}\ and
holds thus a constant probability density everywhere.
So the probability density \mm{P(\rr)}\ is actually measured in
``units of free-electron probability density''.
Thus, by pegging the normalization of  $|\psi_{f}\rangle$\ to the
normalization of the free-electron wave 
function \mm{\exp(\istd \kk \rr)}, one keeps a universal definition
of \mm{P(\rr)}. 

Intuitively, the wave function $|\psi_{\bbox{k}}^{(-)}\rangle$\ can be
viewed as that wave function, from which 
a plane wave evolves within a sufficiently long time.  It represents
the state into which the core electron ``jumps'' as a result of the
electromagnetic perturbation $H_{I}$.  Thus, the quantity $P(\rr)$\ 
 ought to be interpreted as the probability, that the
electron ejected from a core level can be found at position \rr,
``just after'' having absorbed an x-ray photon. 
 By evaluating $P(\rr)$, one
provides the most sensible answer to the seemingly naive question
about the localization of the XANES photoelectron.  We bear in mind, at
the same time, that this Lippman-Schwinger-like description cannot
fully substitute for a proper time-dependent treatment.\cite{breit}

\subsection{Evaluating photoelectron probability density $P(\rr)$}

\label{rovnice}

In this section, we present equations necessary for
calculating the photoelectron probability density $P(\rr)$.  Although
some of them can be found in that or other form in various papers
dealing with \xra\ or \phd\ theory (especially in
Refs.~\onlinecite{nbd,natoli90}), we present them here anyway in
order to embed them into the context of this work, to offer the 
reader a complete set of equations which might be helpful for
practical calculations and, last but not least, to unify various
notations and conventions.

The proper mathematical expression for evaluating the probability
density of electrons participating in \xra\ process can be obtained by
inserting the wave function 
$|\psi_{\bbox{k}}^{(-)}\rangle$\ into Eqs.\ (\ref{pr})--(\ref{wf}).
We get
\begin{equation}
P(\rr) \; = \; 
        \frac{1}{\sigma_{\scriptscriptstyle \text{XAS}}}    \,
        \int \! \dstd^{2}\kkang    \: 
        \frac{\dstd \sigma}{\dstd \Omega_{\bbox{k}}}  \,
      \left| \psi_{\bbox{k}}^{(-)}(\rr) \right|^{2} \; \; ,  
\label{vaha}
\end{equation}
where the PED cross section \mm{\dstd \sigma / \dstd \Omega_{\bbox{k}} }\
stems from the partial probability $w_{f}$\ of Eq.\ (\ref{wf}) and the
XAS cross section 
$\sigma_{\scriptscriptstyle \text{XAS}}$, 
\begin{equation}
\sigma_{\scriptscriptstyle \text{XAS}} \; = \; \int \! \dstd^{2}\kkang 
    \:    \frac{\dstd \sigma}{\dstd \Omega_{\bbox{k}}} \; \; ,
\label{ang}
\end{equation}
ensures correct normalization.

By keeping only the dipole and quadrupole terms in the electromagnetic
hamiltonian $H_{I}$, the PED cross section can be
written as
\begin{eqnarray}
\frac{\dstd \sigma}{\dstd \Omega_{\bbox{k}}} \: = \:
\frac{1}{4 \pi} \alpha\omega k \: 
& & \left[  \, 
 | \langle \psi^{(-)}_{\bbox{k}}| \eeang \rr | \phi_{c}\rangle  |^{2}
 \right. \nonumber \\
& &  + \: \left.
    |\langle \psi^{(-)}_{\bbox{k}}| (\eeang \rr) (\qqang \rr) 
| \phi_{c}\rangle  |^{2} \, \right] \; ,
\label{xas2}
\end{eqnarray}
where \ee\ is the polarization vector of the incoming radiation and \qq\
is its wave vector (\mm{\qq = 1\! /\! 2 \alpha \omega \qqang}).
Employing the muffin-tin approximation,  the wave
functions \psik\ can be expanded inside the $j$-th muffin-tin
sphere as
\begin{equation}
\psik \; = \; \sum_{L} \beta^{(j)}_{L}(\kk) \: 
                       {\cal R}^{(j)}_{\ell}(kr)   \: 
                       Y_{L}(\rrang)  \; \; ,
\label{rozvoj}
\end{equation}
where single-sphere solutions of the radial Schr\"{o}dinger
equation \mm{{\cal R}^{(j)}_{\ell}(kr)}\ are normalized so that they smoothly
match the free-space solution
\begin{equation}
 {\cal R}^{(j)}_{\ell}(kr) \; = \; 
         \cot \delta^{(j)}_{\ell} \, j_{\ell}(kr) 
         \, - \,    n_{\ell}(kr) 
\end{equation}
outside the muffin-tin sphere.  The double-subscript $L$\ stands for
the pair \mm{(\ell,m)}.

Following the formalism of Ref.\ \onlinecite{nbd}, the coefficients 
\mm{\beta^{(j)}_{L}(\kk)}\ can be expanded as 
\begin{equation}
\beta^{(j)}_{L}(\kk) \; = \; 4 \pi \, \sum_{L''}   \istd^{\ell''}   \,
                        \beta^{(j)}_{L}(L'')     \,
                         Y_{L''}^{*}(\kkang) \; \; ,
\label{bk}
\end{equation}
where the amplitudes \mm{\beta^{(j)}_{L}(L'')}\ satisfy
\begin{equation}
\beta^{(j)}_{L}(L'') \; = \; \sum_{p L'}  W^{jp}_{L L'} \,
                                           J^{p0}_{L' L''}
 \; \; .
\label{bl}
\end{equation}
The scattering matrix $W$\ is an inverse matrix to
\widetext
\begin{equation}
\left[ W^{-1} \right]^{ij}_{L L'} \: = \: 
    \left(   \sin \delta^{(j)}_{\ell} \,
             \estd^{-\istd \delta_{\ell}^{(j)}}         \right)^{-1}
   \! \delta_{ij} \delta_{L L'}  \: + \:
    4 \pi\, (1-\delta_{ij})  
    \sum_{L_{1}} \, \istd^{\ell - \ell' + \ell_{1}}
    \: \istd  h^{(-)}_{\ell_{1}}(k|\RR^{ij}|)
    \: Y_{L_{1}}(\RRang^{ij}) 
    \: C^{L'}_{L L_{1}} \; ,
    \label{Winv}
\end{equation}
\narrowtext
where the Gaunt symbol \mm{C^{L'}_{L L_{1}}}\ stands for 
\begin{equation}
C^{L'}_{L L_{1}} \; = \; 
    \int \! \dstd^{2}\nnang \: 
     Y_{L}(\nnang) \,  Y_{L'}^{*}(\nnang)  \,  Y_{L_{1}}(\nnang) 
\; \; ,
\end{equation}
the free-electron propagator \mm{J^{pq}_{L L'}}\ is
\begin{equation}
J^{pq}_{L L'}  \; = \;  
        4 \pi  \sum_{L_{1}} \istd^{\ell - \ell' + \ell_{1}}
        \, j_{\ell_{1}}(k|\RR^{pq}|)
        \: Y_{L_{1}}(\RRang^{pq}) 
        \: C^{L'}_{L L_{1}}
\end{equation}
and $\RR^{ij}$\ is defined as
\begin{equation}
\RR^{ij} \; = \; \RR^{i} - \RR^{j} \; \; .
\end{equation}
The amplitudes \mm{\beta^{(j)}_{L}(L'')}\ and the scattering matrix
\mm{W^{ij}_{L L'}}\ are ``incoming-waves'' analogs of the amplitudes
\mm{B^{(j)}_{L}(L'')}\ and the scattering matrix
\mm{\left[ ( \underline{T} + \underline{H} )^{-1} \right]^{ij}_{LL'}}\
employed in Ref.\ \onlinecite{nbd} for analyzing the scattering of an
electron by a molecule.  See Appendix~\ref{srovnani} for
a more comprehensive comparison.

Employing the amplitudes \mm{\beta^{(j)}_{L}(L'')}, the PED 
cross-section can be expressed via
%
%
%
%
%
%
%
%
%
%
%
%
\begin{eqnarray}
\frac{\dstd \sigma}{\dstd \Omega_{\bbox{k}}}  & = &
            4 \pi \alpha \omega k  \,
            \Biggl\{  \, 
            \left|  
            \sum_{L}  \sum_{L''}  (-\istd)^{\ell^{''}} 
            \left[ \beta^{(0)}_{L}(L'') \right]^{*}
            Y_{L''}(\kkang) \,  D_{L L_{c}} 
            \right|^{2}
            \: +   \nonumber \\
            &  &
            \frac{1}{16} \alpha^{2} \omega^{2} \,
            \left|  
            \sum_{L}  \sum_{L''}  (-\istd)^{\ell^{''}} 
            \left[ \beta^{(0)}_{L}(L'') \right]^{*}
            Y_{L''}(\kkang) \,  Q_{L L_{c}} 
            \right|^{2}
            \,  \Biggr\}
               \; \; ,
\label{phd}
\end{eqnarray}
where the dipole and quadrupole matrix elements are
\begin{equation}
D_{L L_{c}}  \equiv  \int \! \dstd r \, r^{3}  \,
     {\cal R}^{(0)}_{\ell}(kr) \phi_{c}(r) \,
     \int \! \dstd^{2} \rrang \,
     Y_{L}^{*}(\rrang) \, \eeang \rrang \, Y_{L_{c}}(\rrang) 
\end{equation}
and 
\begin{eqnarray}
Q_{L L_{c}} & \equiv & \int \! \dstd r \, r^{4}  \,
     {\cal R}^{(0)}_{\ell}(kr) \phi_{c}(r) \,
     \nonumber \\
 &  & \times \, \int \! \dstd^{2} \rrang \:
     Y_{L}^{*}(\rrang) \, (\eeang \rrang) \, (\qqang \rrang) \,
     Y_{L_{c}}(\rrang)
\end{eqnarray}
and \mm{L_{c}}\ specifies the angular
momentum of the core state located at the central atom~\mm{\RR^{0}}.

The XAS cross section follows from Eq.\ (\ref{ang}) and
Eq.\ (\ref{phd}) as 
\begin{eqnarray}
\sigma_{\scriptscriptstyle \text{XAS}} & = & 4 \pi \alpha \omega k  \, 
            \sum_{L''}  \, \Biggl\{ \, 
            \biggl|   \sum_{L}  
            \left[ \beta^{(0)}_{L}(L'')\right]^{*} \, D_{L L_{c}}  
            \biggr|^{2}   \nonumber \\
           &  & +      
             \frac{1}{16} \alpha^{2} \omega^{2} \,
            \biggl|   \sum_{L}  
            \left[ \beta^{(0)}_{L}(L'')\right]^{*} \, Q_{L L_{c}}
            \biggr|^{2} \, \Biggr\}
               \; ,
\label{xra}
\end{eqnarray}
resembling in this form analogous expressions presented, e.g., in 
Refs.~\onlinecite{dill,nbd} (note that we ignore the electron spin
throughout this paper).  Again, in Appendix~\ref{srovnani} we
will present few other equivalent formulations of Eqs.\ (\ref{phd})
and~(\ref{xra}).

Considering Eqs.\ (\ref{vaha}) and (\ref{rozvoj}), one obtains the
expression for probability density of 
ejected photoelectron inside the $j$-th muffin-tin sphere as
\begin{equation}
P(\rr) \, = \,
        \frac{1}{\sigma_{\scriptscriptstyle \text{XAS}}}    \,
        \int \! \dstd^{2}\kkang    \, 
        \frac{\dstd \sigma}{\dstd \Omega_{\bbox{k}}}  \,
      \biggl| 
       \sum_{L} \beta^{(j)}_{L}(\kk) \,
                       {\cal R}^{(j)}_{\ell}(kr)   \,
                       Y_{L}(\rrang)
      \biggr|^{2}
\label{psph}
\end{equation}
where the PED and XAS cross sections have to be taken from
Eqs.\ (\ref{phd}) and (\ref{xra}).

In electronic structure studies, it might be helpful to have a tool
for investigating the angular-momentum character of photoelectrons
with respect to a site~\mm{\RR^{j}}.  That can be achieved by
inserting into the sum~(\ref{pr}) only the angular-momentum projected
parts of photoelectron wave functions,
\begin{equation}
P_{\ell}(\rr) \; = \; \sum_{f} \, w_{f} \,
            \bigl| {\sf P}_{\ell} \psi_{f}(\rr) \bigr|^{2} \; \; ,
\label{pr_l}
\end{equation}
where \mm{{\sf P}_{\ell}}\ stands for the relevant projection
operator.  Analogously to Eq.\ (\ref{psph}), one gets the probability
density of finding a XANES photoelectron at position \rr\ with an
angular momentum $\ell$\ with respect to the site~\mm{\RR^{j}}\ as 
\begin{equation}
P_{\ell}(\rr) \, = \,
        \frac{1}{\sigma_{\scriptscriptstyle \text{XAS}}}    \,
        \int \! \dstd^{2}\kkang    \, 
        \frac{\dstd \sigma}{\dstd \Omega_{\bbox{k}}}  \,
      \biggl| 
       \sum_{m} \beta^{(j)}_{\ell m}(\kk) \,
                       {\cal R}^{(j)}_{\ell}(kr)   \,
                       Y_{\ell m}(\rrang)
      \biggr|^{2}
\label{psph_l}
\end{equation}
meaning that the sum $\sum_{L}$\ in Eq.\ (\ref{psph}) was just
substituted with $\sum_{m}$.

\subsection{Practical aspects}
\label{sec_practical}

This study is motivated by an effort to understand x-ray absorption
spectra and, in particular, to develop a means of connecting spectral
and structural features.  It is thus desirable to explore how
$P(\rr)$\ depends on the energy of the photoelectron.  
A simple way to extract the fine structure from this dependence 
 is to subtract from $P(\rr)$\  the probability which would
correspond to a single isolated atom --- just like the EXAFS  can be
extracted from a raw absorption 
spectrum by subtracting from it the atomic part (sometimes called
AXAFS or atomic XAFS).\cite{axafs}  The
single-atom 
probability density $P_{\text{ato}}(\rr)$\ can be evaluated 
following the procedure outlined 
 in Sec.~\ref{rovnice}, taking into account only a
single scatterer (cf.\ appendix~\ref{atomic}).
By investigating the difference \mm{P(\rr)-P_{\text{ato}}(\rr)}, one
can see more clearly the effects of surrounding atoms on the formation
of both XANES and the photoelectron \pd.

In practice, one often wants to compare the importance of particular
atoms for generating XAFS.  For that purpose, it
is sufficient to compare not \mm{P(\rr)}\ but rather its integrals
inside suitably chosen 
spheres.  We can define atomic-site-related quantities \mm{P^{(j)}}\
and \mm{\Delta P^{(j)}_{\text{ato}}} 
by 
\begin{mathletters}
\label{jsphere}
\begin{eqnarray}
  P^{(j)} & \equiv & \frac{1}{V^{(j)}} \,
      \int_{0}^{R^{(j)}_{N}} \! \dstd r \, r^{2} \int \! \dstd^{2}\rrang
     \, P(\rr) \; \; ,
     \\
  \Delta P^{(j)}_{\text{ato}} & \equiv & \frac{1}{V^{(j)}} \,
      \int_{0}^{R^{(j)}_{N}} \! \dstd r \, r^{2} \int \! \dstd^{2}\rrang
       \, \left[ P(\rr) - P_{\text{ato}}(\rr) \right]
\end{eqnarray}
\end{mathletters}
where \mm{R^{(j)}_{N}}\ is a suitably chosen normalization radius and
\mm{V^{(j)}}\ is the volume of the normalization sphere around the
$j$-th site.  Note that in the case of muffin-tin approximation, only
spherically-averaged values make sense anyway.

From the XANES analysis point of view, the quantities 
$P^{(j)}$, $\Delta P^{(j)}_{\text{ato}}$\ contain still quite a lot of
unnecessary or ``ballast'' information.  This is due to the fact that
the probability density $P(\rr)$\ defined by
(\ref{vaha}) is dominated by isotropic density of states (DOS)
effects, which are {\em not specific to the site from
which the photoelectron is ejected}.\cite{ja_jpcm}  This DOS-like
contribution can be quantified by defining a DOS-like probability
 density \mm{P_{\scriptscriptstyle \text{DOS}}(\rr)}, which differs
from \Prr\ by assuming a \kkang-independent or ``unidirectional'' \phd\
cross section \mm{\dstd \sigma / \dstd \Omega}\ as\cite{stary}  
\begin{eqnarray}
P_{\scriptscriptstyle \text{DOS}}(\rr) & \equiv & 
      \frac{1}{\sigma_{\scriptscriptstyle \text{XAS}}}    \,
      \frac{\dstd \sigma}{\dstd \Omega}  \,
      \int \! \dstd^{2}\kkang    \, 
      \left| \psi_{\bbox{k}}^{(-)}(\rr) \right|^{2} 
     \nonumber   \\
                 & = &  
     \frac{1}{4 \pi}      \,
     \int \! \dstd^{2}\kkang    \, 
      \left| \psi_{\bbox{k}}^{(-)}(\rr) \right|^{2} \; \; .
 \label{DOS}
\end{eqnarray}
Thus states with different \kkang\ contribute to
\mm{P_{\scriptscriptstyle \text{DOS}}(\rr)}\ with identical weights,
just as is the case of local \rr-dependent DOS.  It can be
shown easily that 
\mm{P_{\scriptscriptstyle \text{DOS}}(\rr)}\ is indeed proportional to the
density of states $n(\rr,E)$, 
\begin{equation}
n(\rr,E) \: = \: - \, \frac{1}{\pi} \, 
    \text{Im} \, G^{(+)}( \rr, \rr ; E) 
         \: = \:
    \frac{k}{4 \pi^2} P_{\scriptscriptstyle \text{DOS}}(\rr) \; \; .
\label{pd_dos}
\end{equation}
An atomic-sphere related quantity 
 \mm{\Delta P^{(j)}_{\scriptscriptstyle \text{DOS}}}\ can be defined
 analogously to Eq.\ (\ref{jsphere}),
\begin{equation}
\label{dos_sphere}
  \Delta P^{(j)}_{\scriptscriptstyle \text{DOS}} \equiv 
 \frac{1}{V^{(j)}} \,
      \int_{0}^{R^{(j)}_{N}} \! \dstd r \, r^{2} \int \dstd^{2}\rrang
       \, \left[ P(\rr) - P_{\scriptscriptstyle \text{DOS}}(\rr) \right]
      \; \; .
\end{equation}

Typically, $P$\ does not differ from $P_{\scriptscriptstyle
\text{DOS}}$\ by more than 10\% (but often much less). Specific
XAS-related effects may thus be obscured in $P(\rr)$\ by more general
DOS effects.  The difference probability density, \mm{
P(\rr)-P_{\scriptscriptstyle \text{DOS}}(\rr) }, informs how the
spatial localization of a XANES electron differs from the spatial
localization of a ``generic'' electron (with the same energy).  By
investigating the difference probability \mm{ \Delta
P^{(j)}_{\scriptscriptstyle \text{DOS}}}, one can filter out effects
which are not specific for XAS.

Generally, the total \pd\ \pj\ informs where the most of the
photoelectron is located, while the difference \pd\ \dpjdos\ is
especially sensitive to the photon polarization \ee\ and to the
position of the photoabsorbing atom.  We postpone a further discussion
of these concepts to Sec.\ \ref{sec_tis2}, where they will be
illustrated on a concrete example.

\section{Application of photoelectron probability density analysis}

\label{aplikace}

We explore the potential of photoelectron probability density (PEPD)
analysis by examining several XANES spectra, which were subdued to
investigation of the origin of their peaks in the past.  In
particular, we will concentrate on the pre-peak at Ti {\it K} edge
XANES of \tis\ and on the whole Ti \kedge\ XANES of rutile \tio.

All the calculations presented here were done for a non
self-consistent muffin-tin potential constructed via Mattheiss
prescription (superposition of charge densities of isolated atoms).
Using non self-consistent potentials is not a serious drawback in this
case as our aim is not to achieve the best reproduction of
experimental spectra but to demonstrate a method how the calculated
spectra can be analyzed.

 The exchange-correlation potential of Ceperley and
Adler\cite{pickett} was used for atomic calculations of occupied
states.  In constructing the Mattheiss potential appropriate for
unoccupied states, an energy-independent $X\alpha$\ potential with the
Kohn-Sham value of $\alpha=0.66$\ was used.\cite{kohn} Structural data
were taken from the {\sc crystin} database.\cite{crystin} Only dipole
transitions were taken into account.  This limitation is justified in
this study because it was demonstrated experimentally that quadrupole
transitions do not contribute to the \tis\ pre-preak intensity
significantly\cite{tix2} and in the case of \tio\ spectrum our concern
is with the main peaks and the extended XANES region, where the
quadrupole contribution again is negligible.  Muffin-tin radii of
nonoverlapping spheres were determined so that single-site potentials,
which were being superimposed, matched at the touching points
(``matching potential condition'').  The muffin-tin zero was set to
the average interstitial potential.  The influence of the core hole
left on the central atom by the excited electron was taken into
account by calculating the central atom charge density with one
electron being moved from the 1$s$\ core level to the lowest
unoccupied atomic orbital (a relaxed and screen model).  A more
thorough discussion of how the potential is constructed can be found,
e.g., in Refs.\ \onlinecite{v2o5,cugase2}.

When evaluating the sphere-averaged quantities \pj, \dpjat, and
\dpjdos\ according to Eqs.\ (\ref{jsphere}) and (\ref{dos_sphere}),
one has to choose the normalization radius \rn\ for each of the
spheres.  Throughout this section, we always take \rn\ identical for
all atoms of a given compound and equate it with the smallest of
muffin-tin radii.  We found that changing \rn\ affects rather the
overall magnitude of PEPD than its fine structure. Consequently, the
overall picture, as presented in Sec.~\ref{sec_tis2}--\ref{sec_tio2},
does not depend on the choice of \rn.

Experimental \xra\ spectra are broadened with respect to ``raw''
theoretical spectra because of various many-body processes (finite
core hole lifetime, extrinsic photoelectron losses etc.).  It is thus
desirable to anticipate this smearing when evaluating the PEPD's ---
otherwise, one would be overburdened with too many details with little
or no physical significance.  Therefore, we modified the scattering
potential by adding to it a small negative imaginary part.  We set its
magnitude so that it simulates lorenzian broadening equivalent to the
half of a natural core hole width (taken from compilation of Al~Shamma
\ea).\cite{lifetime} Such a procedure would certainly be insufficient
if one tried to describe the spectral broadening in a realistic way
--- apart from too small core hole smearing, energy-dependent
inelastic energy losses of the photoelectron are unaccounted for
altogether.  However, as our main aim is to analyze the peaks in the
{\em calculated} spectrum, we prefer to include less damping in order
to have certain features more pronounced.  We checked that increasing
the imaginary potential twice would not affect the outcome of our
analysis considerably.

\subsection{Pre-peak at Ti \kedge\ of \tis}

\label{sec_tis2}

\tis\ has a layered crystal structure consisting of
sulphur-titanium-sulphur slabs.  Each of these slabs is formed by a
twodimensional hexagonal titanium sublattice sandwiched by two closely
adjacent sulphur hexagonal sublattices.  Locally, the titanium atom is
octahedrally coordinated by six sulphur atoms (cf.\
figure~3 of Ref.\ \onlinecite{wu}),  the  second coordination
shell is formed by Ti atoms of the same  hexagonal  sublattice to
which the central Ti atom belongs.

The polarized Ti \kedge\ XANES of \tis\ shows a
distinct pre-peak at the $xy$\ polarization (i.e., when the
polarization vector of the incoming radiation lies within the titanium
layer).\cite{wu,tonda}  Relying on XANES calculations for
artificial trial structures, Wu \ea\cite{wu} suggested that the
pre-edge is generated by multiple scattering which involves mainly the
central Ti atom and the Ti atoms of the second coordination
shell.

%
%

\begin{figure}
\includegraphics[0mm,-5mm][165mm,125mm]{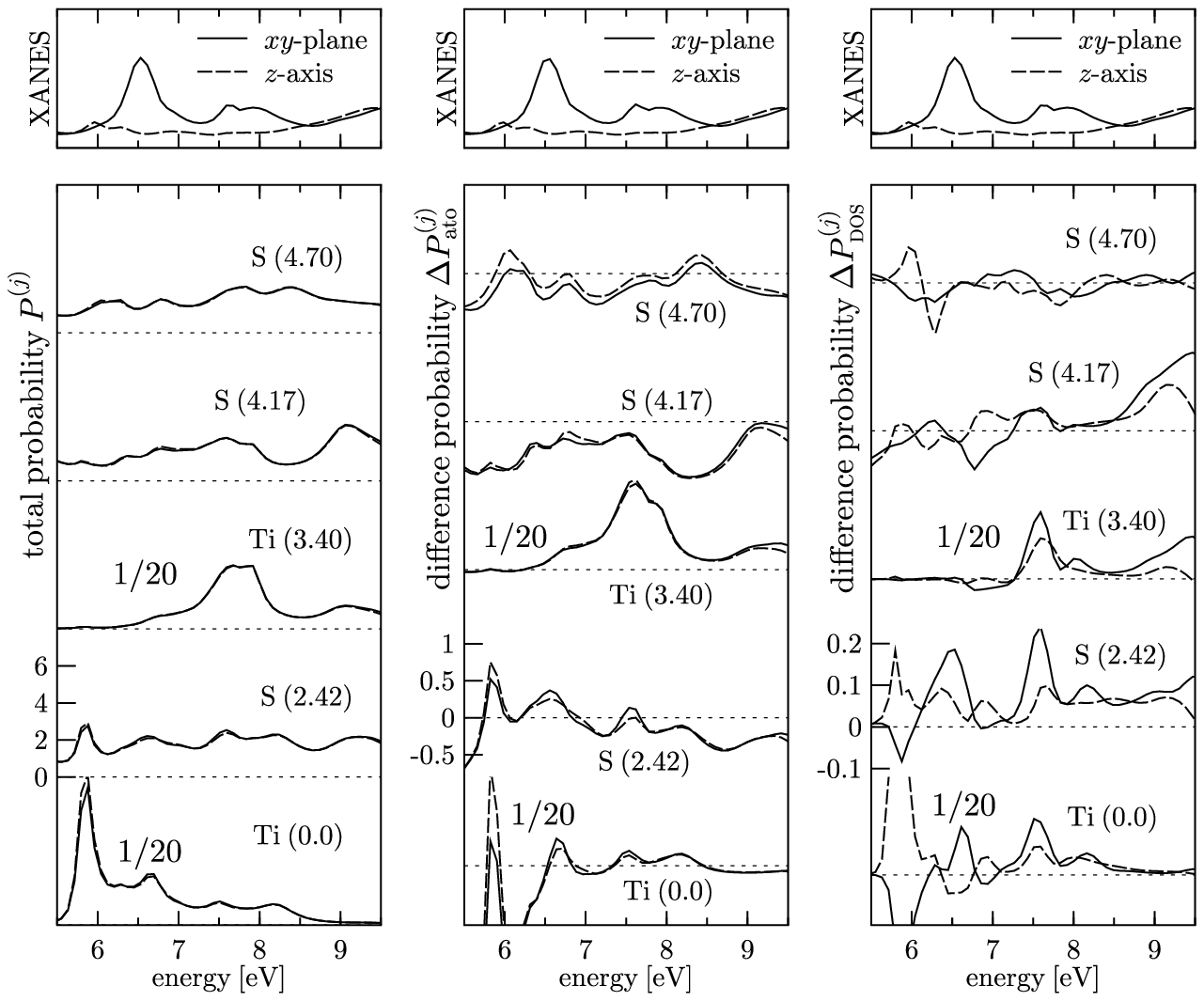}
\caption{Theoretical pre-peak structure of polarized Ti \kedge\ XANES
of \tis\ (upper panels), together with the photoelectron \pd\ \pj\
(lower left panel), 
atomic difference \pd\ \dpjat\ (lower middle panel), and DOS
difference \pd\ \dpjdos\ (lower right panel) around selected atoms.
Solid lines 
correspond to the \exy\ polarization, dashed lines to the \ez\
polarization.    PEPD curves are
identified by the 
chemical type of the appropriate atom and by its distance in~\AA\ from
the center of the 135-atoms cluster. 
Thin dotted lines mark zeros of \pj, \dpjat, and \dpjdos\ for
each subgraph. The absolute scale of the PEPD's is indicated at 
the lower panels, the scale of XANES is 
arbitrary.  Note that curves for Ti atoms were divided by 20,  
as indicated.}
\label{fig_tis2}
\end{figure}

A comprehensive analysis of the Ti \kedge\ XANES of \tis, including
comparison between theory and experiment, can be found
elsewhere.\cite{wu,tonda,tix2}  Therefore, we present in
Fig.\ \ref{fig_tis2} only the theoretical polarized pre-peak structure,
calculated for a cluster of 135 atoms, 
together with \pj, \dpjat, and \dpjdos\ curves for the central atom
and for atoms 
of its four nearest coordination shells.  Two  polarizations 
correspond to the \ee\ vector averaged over the full $2\pi$\
angle within the $xy$\ plane and to the \ez\ setup.
Titanium probability densities are scaled down by a factor of 
20 with respect to corresponding sulphur curves, as indicated.
Individual atoms belonging to the same  
coordination shell give rise to identical PEPD's in this case
(also for further shells than those displayed in Fig.\ \ref{fig_tis2}).
As we did not perform a band-structure calculation, the Fermi level is
not fixed.  It follows from Fig.\ \ref{fig_tis2} that it
ought to be around \mm{\approx 6}~eV above the muffin-tin zero,
therefore, all states bellow it are actually occupied and do not
contribute to \xra\ spectrum --- we show them just for completeness. 

The first information one gets from Fig.\ \ref{fig_tis2} is that the
photoelectron probability density clearly oscillates with energy.  The
fine structure in PEPD differs from the XAFS.  One might be surprised
at first by the fact that this is the case not only for non-central
atoms but also for the photoabsorbing one (after all, XAS is
approximately proportional to unoccupied DOS angularly projected on
the central atom).  One has to recall, however, that PEPD and XAS
carry in fact different kind of information:   While the XANES intensity
informs us about the probability that the ejected electron goes
anywhere, PEPD tells us how this ``anywhere'' looks like.  Only a tiny
fraction of the photoelectron density matrix (\ref{denmat}), namely,
that part of it which overlaps with the core of the central atom and
which has the angular-momentum character conforming to the dipole
and/or quadrupole selection rules, enters indirectly into expressions
for the XANES intensity (\ref{signbd})--(\ref{sigtau}).

One can see immediately from the lower left panel of Fig.\
\ref{fig_tis2} that the chemical type governs the gross shape of \pj\
curves (we checked that this is true also for more distant atoms,
which are not displayed here).  The differences between PEPD's around
atoms at crystallographically equivalent positions (say, Ti or S atoms
belonging to different coordination spheres) comes first of all from
the fact that this quantity is specifically related to the position of
photoabsorbing atom, i.e., the site from which the photoelectron
was ejected.  (Apart from that, our finite-cluster approach obviously
introduces inequality among otherwise equivalent sites.)

Hardly any polarization effect can be noticed in the \mm{P^{(j)}}\ or
$\Delta P^{(j)}_{\text{ato}}$\ curves, despite the fact that the
\ee-dependence in the XANES spectrum is quite significant.  Only the
difference probability \dpjdos, which emphasizes the effect of the
particular site from which the photoelectron is ejected, displays a
strong polarization dependence.  This is a manifestation of the
dominance of the DOS-generated unidirectional
\mm{P_{\scriptscriptstyle \text{DOS}}^{(j)}}\ contribution to the
total PEPD, as mentioned in the end of Sec.\ \ref{sec_practical}.

Neither local DOS nor
$P_{\scriptscriptstyle \text{DOS}}$\  take into account that the
photoelectron is {\em ejected from a particular site} via a dipole
transition\cite{ja_jpcm} --- they are concerned with all electrons
of a given energy equally. 
On the other hand, the probability \pj\ that a Ti \kedge\ XAS photoelectron
will be found near atom $j$\ differs, albeit slightly, from the
probability of finding there ``any'' electron of the same energy.
It is this small 
difference which reflects the fact that the \tis\ crystal does not
look identical when viewed from the Ti site either in the $xy$-plane
or in the $z$-axis directions.  So one has to resort to the difference
probability \dpjdos\ if
the polarization-related features of PEPD are to be
studied.

On the other hand, simple
subtraction of the single-atom \pd\ \mm{P^{(j)}_{\text{ato}}}\ does
not provide a new insight.  Due to the 
smoothness of atomic \pd\ $P_{\text{ato}}^{(j)}$, the  $P^{(j)}$\ and 
\dpjat\ curves look very similar, as can be seen in the left
and middle panels of Fig.\ \ref{fig_tis2}.  The total \pd\ \pj\ and
atomic difference 
\pd\ \dpjat\ carry essentially identical pieces of information.

As it follows from Fig.\ \ref{fig_tis2}, there seems to be 
 no simple correspondence between XANES peaks and peaks in the
photoelectron \pd.  Although for some atoms and/or peaks,
one can establish a visual connection between XANES and PEPD, 
 for other features such a discernible connection is clearly
absent.  This lack of simple correspondence between XANES and PEPD
peaks may be a manifestation
of the interference nature of XAFS ---
it is generated not just through accumulating  electrons here and
there but rather by interference between many scattering paths.

We performed also an angular-momentum analysis of PEPD, according to
Eqs.\ (\ref{pr_l})--(\ref{psph_l}).  We found that the $d$\ component
dominates at Ti atoms (it contributes by more that 95\% to either \pj\
or \dpjat\ or \dpjdos) and that the $p$\ component prevails at S atoms
(comprises 60--70\% of PEPD).  Moreover, practically all of the fine
structure in PEPD is formed by the dominant $\ell$-component (i.e.,
$d$\ at Ti and $p$\ at S atoms).  This might again look surprising for
someone who is accustomed to the conventional slang that the excited
photoelectron has a $p$\ character at the \kedge\ due to the dipole
selection rule.  The point is that this would apply literally only
in case that the final photoelectron state were an eigenstate of the
angular momentum.  On the other hand, the wave function
$|\psi_{\bbox{k}}^{(-)}\rangle$, which describes the exited
photoelectron (as argued in Sec.\ \ref{sec_wave}), is not an
angular-momentum eigenstate.  Rather, it is a superposition of states
with different angular momenta, and the component with $\ell$=2
dominates at the central Ti site as a whole, while the relatively tiny
component with $\ell$=1 determines the XANES intensity.

As mentioned, Wu \ea\cite{wu} suggest that the second-shell Ti atoms
play a crucial role in generating the \exy\ pre-peak.  They arrived at
this conclusion by observing that this pre-peak disappears if those
six Ti atoms are removed from the cluster.  A closer look at Fig.\
\ref{fig_tis2} reveals that the DOS-inclusive probability \pj\ is,
indeed, much higher near Ti atoms than near S atoms.  However, this
effect is clearly DOS-related, having little connection with the
particular direction of the photoelectron trajectory.  By inspecting
the DOS-corrected difference \pd\ \dpjdos\ corresponding to the
dominant pre-edge peak at $E$=6.5~eV, we can see that the largest
effects {\em purely connected with \xra}\ actually occur at the
nearest sulphur atoms.  It is thus clear that nearest sulphurs
definitely have their role in generating the \exy\ pre-peak.  In fact,
we found that omitting six nearest S atoms from a large 135-atoms
cluster changes the calculated XANES drastically (making the very
concept of pre-edge region inapplicable).  Only the
secondary-in-importance pre-edge peak at \mm{E \approx}7.5--8.0~eV
seems to be generated by scattering off second-shell Ti atoms, as it
follows from \dpjdos\ curves in Fig.\ \ref{fig_tis2}.  It seems,
therefore, that the physical picture of the process which is
responsible for the \exy\ pre-peak should consider the nearest sulphur
atoms into account.  Namely, although the total amount of the
photoelectron density \pj\ is small on sulphurs as compared with
titaniums, its relatively modest variations with energy and
polarization have a big impact on the XANES.

Let us emphasize that, strictly speaking, this kind of analysis
only informs 
about sensitivity of the photoelectron density near individual atoms
to the changes of polarization vector direction.  It does not testify
about the physical mechanism which may stand behind the
creation of the pre-peak electron states.  So our conclusions do
not in fact contradict the suggestions that those states arise due to
hybridization of central Ti 4$p$\ and next-neighboring Ti 3$d$
orbitals\cite{wu} --- both views may be rather complementary than
opposing.

Finally, let us note that the new look on the role of nearest sulphurs
we take here may be relevant to other non-centrosymmetric systems
with a distinct XANES pre-peak, such as a colossal magnetoresistence 
material La$_{1-x}$Ca$_{x}$MnO$_{3}$\ (Ref.\ \onlinecite{mno}).

\subsection{Ti \kedge\ XANES of rutile \tio}

\label{sec_tio2}

The Ti \kedge\ XANES of rutile \tio\ was studied very intensively due
to its interesting pre-edge structure and the debate still does not
seem to be settled.\cite{farges,brouder90,uozumi,wu_tio,joly}
In this paper, however, we want to concentrate on the extended XANES
 up to $\sim$120~eV, as a comprehensive study of the effect of
individual atoms on spectral peaks in this
region was performed.\cite{jeanne1,jeanne2}  In rutile \tio, the Ti atoms are
located at the center of a distorted octahedron. The plane containing
four equatorial oxygens \oeq\ is parallel to the crystallographic $c$\
axis while two  axial oxygens \oax\ lie inside the plane defined by
axes $a$\ and $b$.  The third coordination sphere is formed by
two titanium atoms \tic, shifted from the central Ti by 
\mm{\pm \bbox{c}}.  Instructive depictions of rutile \tio\
structure can be found, e.g., in figure~3 of
Ref.\ \onlinecite{wu_tio}, figure~2 of Ref.\ \onlinecite{jeanne2}, 
or figure~3 of Ref.\ \onlinecite{poumel_obr}. 

Jeanne-Rose and Poumellec\cite{jeanne2} studied how the polarized
Ti \kedge\ XANES of \tio, calculated for a cluster of 25 atoms, is
affected by variations in the positions of \oax\ and \oeq\ atoms.  
That makes it possible for us to explore here to what
extent does sensitivity of a spectral peak to a position of a certain
atom imply high PEPD around that atom and {\em vice
versa}. In accordance with the discussion in Sec.\ \ref{sec_tis2}, we
will concentrate more on the difference probability \dpjdos\ than on
\pj\ itself.

%
%

\begin{figure}
\includegraphics[25mm,-10mm][175mm,150mm]{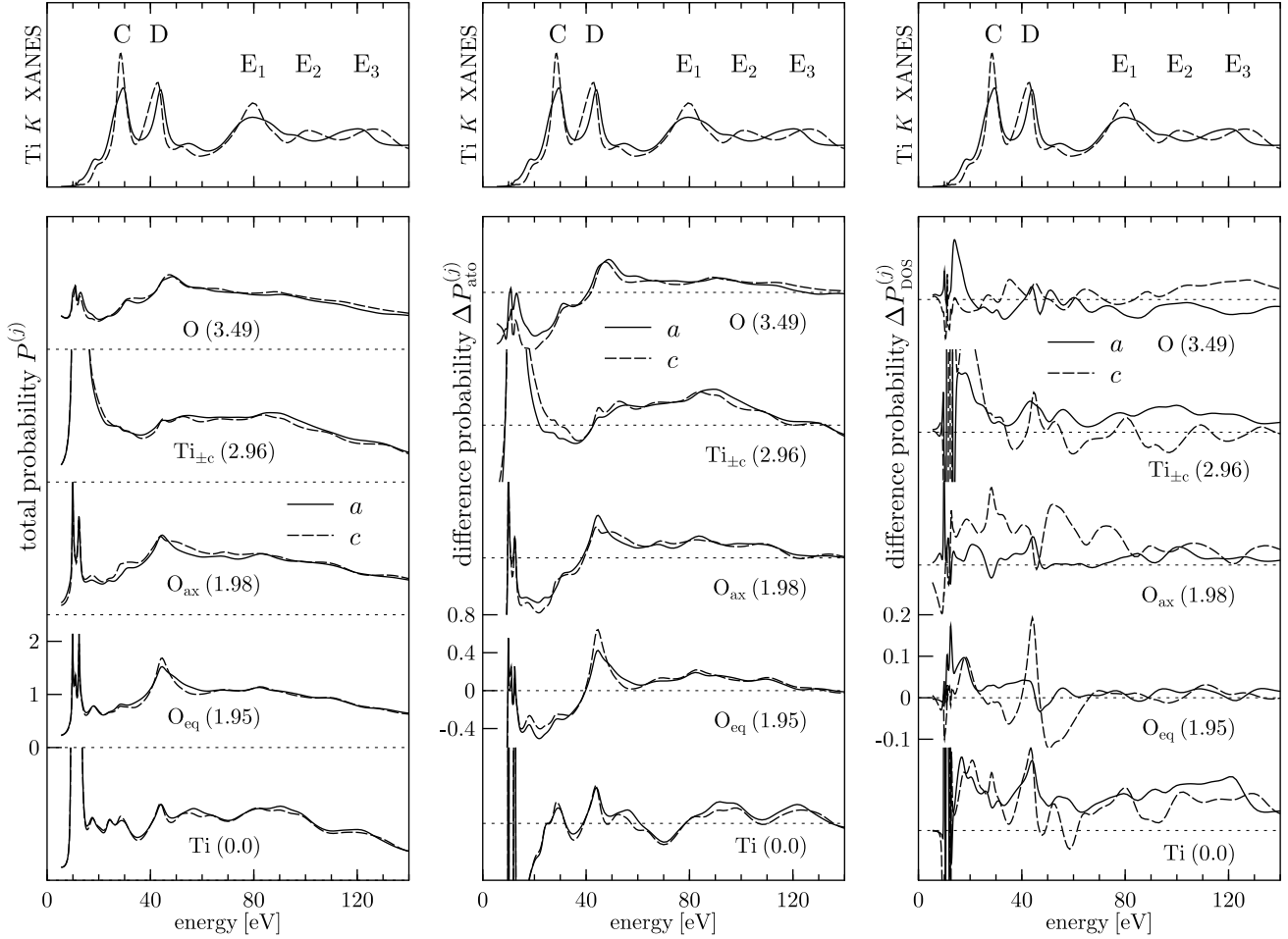}
\caption{Theoretical  polarized Ti \kedge\ XANES
of \tio\ and corresponding PEPD curves.  The meaning of
the curves and symbols is analogous as in 
Fig.\ \protect\ref{fig_tis2}.}
\label{fig_tio2}
\end{figure}

  In Fig.\ \ref{fig_tio2}, theoretical polarized Ti \kedge\
XANES of \tio\ is shown together with corresponding \pj, \dpjat, and
\dpjdos\ curves for the central Ti and its nearest neighbors.  In
order to connect with 
Jeanne-Rose and Poumellec,\cite{jeanne2} we consider a cluster of 25
atoms. The notation of spectral peaks 
$C$, $D$, $E_{1}$, $E_{2}$\ and $E_{3}$\ is taken from
Ref.\ \onlinecite{jeanne2}, too.  The {\em ad-hoc} Fermi energy would
be around 10~eV in the energy scale of Fig.\ \ref{fig_tio2}, meaning that the
wild oscillations at the beginning of the PEPD curves actually fall
predominantly into the region of occupied states.

Let us now compare the conclusions of Jeanne-Rose and
Poumellec\cite{jeanne2} with the picture offered by
the PEPD analysis. 
The equatorial oxygen \oeq\ was found to influence both the C and the
D peaks for both polarizations, the effect being significantly 
stronger on D than on C and for the \ecc\ polarization than for the
\eaa\ polarization.\cite{jeanne2}  A counterpart to this effect can be
found in the PEPD:   A distinct peak in \dpjdos\ around \oeq\
can be found at D and a mild feature at C for both polarizations; the
D peak in \dpjdos\ is higher for the \ecc\ than for the \eaa\ polarization.

When the position of
the axial oxygen \oax\ varies, the calculated \eaa\ XANES is drastically
altered at the D peak and not so much at the C peak, while the \ecc\
spectrum is 
changed at the C peak only.\cite{jeanne2}  A brief look at
the \oax\ curves in Fig.\ \ref{fig_tio2} reveals
that the largest \dpjdos\ around the \oax\ atom is at the D peak for the
\eaa\ spectrum and at the C peak for the \ecc\ case.  So indeed, 
the energy at which a particular atom affects the photoelectron
probability density most prominently coincides in this case with the
energy at which the XANES is  
significantly changed when that atom is moved.  The analogy is
nevertheless only a qualitative one --- larger 
\dpjdos\ does not necessarily imply bigger changes of XANES
peaks when \oax\ is moved. For example, the C resonance in \dpjdos\ 
 for \ecc\ is stronger than the D resonance for
\eaa, and yet the exact position of the \oax\ atom affects more the 
D spectral peak at \ecc\ than the 
C peak at \eaa\ polarization.\cite{jeanne2}
On the other hand, the total \pd \pj\ is larger at the D peak energy
than at the C peak for both oxygens, reflecting correctly the greater 
sensitivity of the D peak to their positions. 

 So the following intuitively plausible picture emerges:  For
 comparing XANES peaks at different energies, the total probability
 \pj\ may be a good indication of their respective sensitivity to atomic
 positions. For investigating polarization effects, one has to resort
 to \dpjdos.

Unlike for the C and D peaks, a correspondence between XANES and
 \dpjdos\ peaks  cannot be established for the E$_{i}$\ 
 maxima.  Jeanne-Rose and Poumellec\cite{jeanne2} found that 
the E$_{2}$\ peak in the \ecc\ spectrum is
affected quite a lot by the \oeq\ movement while peaks E$_{1}$\ and
E$_{3}$\ are left basically intact for either
polarization -- a property that does not seem to have a counterpart in
 Fig.\ \ref{fig_tio2}.  The sensibility of a XANES peak to
a movement of a particular atom, hence, does not necessarily imply a high
 localization of the photoelectron around that atom --- that
 can be found only by a proper PEPD calculation.

It is worth noting that if single-scattering dominates in generating
a particular XANES peak (such as C for the \eaa\
polarization),\cite{jeanne2} one can observe a distinct resonance in
\dpjdos\ at corresponding energy.  On the other hand, such a
correspondence appears to be blurred for spectral peaks where a significant
contribution from multiple-scattering is suspected (such as C peak 
 at the \oax\ atom for the \eaa\ polarization or the E$_{2}$\ peak at \oeq\
and \tic\  for the \ecc\ polarization).\cite{jeanne2}  This seems to
be plausible --- 
a multiple-scattering nature of a peak emphasizes that it is generated not
by a mere ``presence'' of the photoelectron near certain atom but rather
by a complicated interference process.
It remains to be explored to what extent this trend is a general one.

\section{Discussion}

Photoelectron probability density offers us a direct look on the
spatial localization of those electron states which are seen by XAS.
Contrary to a bit vague concepts like importance of various atoms for
the formation of a particular XANES peak or order of
multiple-scattering which has to be accounted for, PEPD is a
rigorously defined quantity with a transparent physical
interpretation.  Hence even if there were no immediate practical
applications, PEPD analysis would still remain a valuable tool
for understanding the physical processes which give rise to the
 XANES spectrum.  Apart from this principal asset, there are two
directions where PEPD analysis could contribute to solving concrete
problems:  investigations of real structure of solids and
investigations of their electronic structure.

One of the obstacles to overcome when fitting the experimental XANES 
spectrum with calculated spectra of trial structures is that ones has,
in general, a lot of atoms to move and hence it is difficult to
identify those whose positions are most critical for the XANES shape.
By providing a deeper insight into the 
``photoelectron trajectory'', PEPD analysis may drop 
 a hint for the most critical spots in advance.
Intuitive arguments and practical experience (Sec.\
\ref{aplikace}) show that both \pj\ and \dpjdos\ reflect the
sensitivity of XANES to movement of individual atoms:  Total \pd\ \pj\
is more relevant for comparing roles of different chemical species
while \dpjdos\ is more indicative of the polarization and
site dependence of XANES spectra. 

In electronic structure studies, the appealing feature of PEPD is that
it directly investigates those unoccupied states which are probed by
XAS.  When investigating the spatial localization of these states, one
is thus not left relying on heuristic arguments like ``which atom
affects the spectrum most if moved''.  One can even explore the
angular-momentum character of XANES states by projecting the wave
functions \mm{\psi^{(-)}_{\bbox{k}}}\ along
 Eqs.\ (\ref{pr_l})--(\ref{psph_l}).  In that way, one can see not
only from which atom a particular XANES peak arises but also from what
type of orbital is comes from.

The concept of PEPD relates, of course, not just to the XANES region
but to the EXAFS part of the absorption spectrum as well. The reason
why we mention only XANES here explicitly is that EXAFS oscillations
can be properly analyzed with other tools.  Note also that the PEPD
analysis could be applied to photoelectron diffraction as well --- one
would just have to omitt the angular integration 
\mm{\int \! \dstd^{2}\kkang}\ in Eqs.\ (\ref{vaha}), (\ref{psph}) and
in related expressions.

\section{Conclusions}

Our analysis demonstrates that it makes sense to explore
the probability that a photoelectron participating in a XANES
process can be found at a specific place.  The relevant quantity
is the photoelectron probability density and it can be 
calculated as a sum of squares of wave functions describing elementary
PED processes, weighted by normalized PED cross sections.  
When investigated as a function of the photoelectron energy, it
exhibits a resonance-like structure and depends  
on  the atom around which it is evaluated.
The bulk of PEPD is dominanted by DOS-like effects, meaning that,
e.g.,  hardly any polarization dependence can be seen in it unless
the DOS-related portion is subtracted.
In many cases, high difference probability \dpjdos\ around an atom may
serve as an indication of high sensitivity of XANES towards the
position of that atom for a given energy.
The fine structure in PEPD does not copy the corresponding XANES:
For some atoms and/or peaks, a visual  connection between XANES
and PEPD can be established,  for other features such a
discernible connection is absent.  This may be a manifestation
of the interference nature of XAFS ---
the correspondence between XAS and PEPD peaks is more often 
observed for features which arise from single scattering than for
multiple-scattering peaks.
The spatial dependence of PEPD thus provides information
which is not equivalent to what can be learned from comparing 
theoretical spectra for various trial structures but rather is
complementary to it.

By performing a PEPD analysis for a Ti \kedge\ of \tis, we found that,
contrary to earlier interpretations, the sulphur atom nearest to the
absorbing titanium participates significantly in formation of the
distinct polarization-dependence of the pre-peak.

\section*{Acknowledgements}

This work was supported by the grant 202/99/0404 of the Grant Agency of
the Czech Republic.  The use of the {\sc crystin} structural
database was financed by the grant 203/99/0067 of the Grant Agency of
the Czech Republic.  The author is grateful to A.\
\v{S}im{\accent23 u}nek for stimulating discussions and for a critical
reading of the manuscript.

\appendix

\section{Other expressions\\
 for PED and XAS cross sections} 

\label{srovnani}

In this appendix, some other formulations of the equations presented
in Sec.~\ref{rovnice} are shown, in order to facilitate connection
with notations and definitions in other works.

The \phd\ cross section, evaluated in Eq.\ (\ref{phd}), can be
expressed as
\begin{eqnarray}
\frac{\dstd \sigma}{\dstd \Omega_{\bbox{k}}}  & = &
            \frac{1}{4 \pi} \alpha \omega k  \,
            \Biggl[  \, 
            \biggl|  
            \sum_{L}  \beta^{(0)}_{L}(\kk) D_{L L_{c}}^{*} 
            \biggr|^{2}  \nonumber \\
            & & + \: 
            \frac{1}{16} \alpha^{2} \omega^{2} \,
            \biggl|  
            \sum_{L}  \beta^{(0)}_{L}(\kk) Q_{L L_{c}}^{*} 
            \biggr|^{2}
            \,  \Biggr]
               \; \; .
\label{phd_1}
\end{eqnarray}
By introducing the scattering matrix $W$\ of Eq.\ (\ref{Winv}) directly into
Eq.\ (\ref{phd}), one obtains 
%
%
%
%
%
%
%
%
%
%
%
%
\begin{eqnarray}
\frac{\dstd \sigma}{\dstd \Omega_{\bbox{k}}}  & = &
            4 \pi \alpha \omega k  \,
            \Biggl[  \, 
            \left|  \sum_{L} \sum_{j L''}
            W^{0j}_{L L''} \istd^{\ell''} Y_{L''}^{*}(\kkang)
            \estd^{\istd \bbox{k} ( \bbox{R}^{j} - \bbox{R}^{0} ) }
            D^{*}_{L L_{c}}  \right|^{2} \; + \nonumber \\
            & &
            \frac{1}{16} \alpha^{2} \omega^{2} \,
            \left|  \sum_{L} \sum_{j L''}
            W^{0j}_{L L''} \istd^{\ell''} Y_{L''}^{*}(\kkang)
            \estd^{\istd \bbox{k} ( \bbox{R}^{j} - \bbox{R}^{0} ) }
            Q^{*}_{L L_{c}}  \right|^{2}
            \Biggr]
               \; \; ,
\label{phd_2}
\end{eqnarray}
which is analogous to equation~(22a) of Natoli
\ea.\cite{natoli90}

As a lot of useful relations about XAS and PED within the {\em wave
function formalism} can be found in Ref.\ \onlinecite{nbd}, we quote
here the relation between the ``outgoing'' quantities used in that
paper and the ``incoming'' quantities employed here.  The
scattering amplitude \mm{\beta^{(j)}_{L}(L'')}\ of Eq.\ (\ref{bl}) is
connected with the amplitude \mm{B^{(j)}_{L}(L'')}\ of equation (2.28)
of Natoli \ea\cite{nbd} via
\begin{equation}
\beta^{(j)}_{L}(L'') \; = \;
        (-1)^{m} \, \left[
        B^{(j)}_{\ell,-m}(\ell'',-m'')
        \right]^{*}    (-1)^{m''} 
\end{equation}
and the scattering matrix $W$\ relates to the matrix 
\begin{displaymath}
Z^{ij}_{LL'} \equiv 
    \left[ ( \underline{T} + \underline{H} )^{-1} \right]^{ij}_{LL'}
\end{displaymath}
 of equation (3.3) of Natoli \ea\cite{nbd} as
\begin{equation}
 W^{ij}_{L L'} \; = \; 
             (-1)^{m}  \,  \left[
             Z^{ij}_{(\ell,-m)(\ell',-m')} 
             \right]^{*}  \,  (-1)^{m'} \; \; ,
\end{equation}
and to the $\tau$\ matrix of Refs.\ \onlinecite{dph,vsp} as
\begin{equation}
 W^{ij}_{L L'} \; = \; 
             - \, (-1)^{m'}  \,  \left[
             \tau^{ji}_{(\ell',-m')(\ell,-m)} 
             \right]^{*}  \,  (-1)^{m}  
       \; \; .
\end{equation}
Employing the \mm{Z}\ and $\tau$\
matrices, the \xra\ cross section of Eq.\ (\ref{xra}) can be written 
as 
\begin{eqnarray}
\sigma_{\scriptscriptstyle \text{XAS}} & = & 4 \pi \alpha \omega k \,
    \sum_{L L'} \Bigl\{  \,
     \text{Im} \left(
     D^{*}_{L L_{c}} 
     Z^{00}_{L L'}
     D_{L' L_{c}}  \right) \nonumber \\
  & &  + \: 
    \frac{1}{16} \alpha^{2} \omega^{2} 
     \text{Im} \left(
     Q^{*}_{L L_{c}} 
     Z^{00}_{L L'}
     Q_{L' L_{c}}   \right) \, 
     \Bigr\}
\label{signbd}
\end{eqnarray}
or as 
\begin{eqnarray}
\sigma_{\scriptscriptstyle \text{XAS}} & = & - \, 4 \pi \alpha \omega k \,
    \sum_{L L'} \Bigl\{  \, 
     \text{Im} \left(
     D_{L L_{c}} 
     \tau^{00}_{L L'}
     D^{*}_{L' L_{c}}     \right) \nonumber \\
    & & + \:
    \frac{1}{16} \alpha^{2} \omega^{2} 
     \text{Im} \left(
     Q_{L L_{c}} 
     \tau^{00}_{L L'}
     Q^{*}_{L' L_{c}}               \right) \, 
     \Bigr\} \; \; .
\label{sigtau}
\end{eqnarray}
Eqs.\ (\ref{signbd})--(\ref{sigtau}) can be arrived at either by
invoking the Green function formalism from the beginning or by
applying the optical theorem on Eq.\ (\ref{xra}), as outlined by Natoli
\ea.\cite{nbd}

\section{\\ PEPD in case of a single atom}
\label{atomic}

In calculating the single-atom
probability density $P_{\text{ato}}(\rr)$, one can proceed just as
in Sec.~\ref{rovnice}.  The only difficulty arises in expanding the 
solution of a single-atom Lippman-Schwinger equation inside (now)
empty spheres around sites $\RR^{j}$.  One cannot 
mechanically employ equations (\ref{bk})--(\ref{Winv}) to find the
amplitudes \mm{\beta^{(j)}_{L}(\kk)}, as the phaseshifts at
non-central atomic sites are zero now.  Instead, by applying formula
for re-expanding 
\mm{j_{\ell}(k|\RR^{p}|)\, Y_{L}(\RR^{p}) }\ and 
\mm{n_{\ell}(k|\RR^{p}|)\, Y_{L}(\RR^{p}) }\
around different origins, one can arrive at the following expression
for \mm{\beta^{(j)}_{L}(L'')}, 
\begin{eqnarray}
\beta^{(j)}_{L}(L'') & = &
        \estd^{-\istd \delta_{\ell''}^{(0)}} \,
          4 \pi  \sum_{p L'} 
       \istd^{\ell - \ell'' + \ell'} \,
        \left[ 
        j_{\ell'}(k|\RR^{p}|) \cos\delta^{(0)}_{\ell''}
       \right. \nonumber \\
   & &   \left.
      - \, n_{\ell'}(k|\RR^{p}|) \sin\delta^{(0)}_{\ell''}
        \right] \,
        Y_{L'}(\RR^{p}) 
        \, C^{L''}_{L L'}
      \; \; .
\end{eqnarray}
which ought to be inserted into Eq.\ (\ref{psph}) with the help of
(\ref{bk}), together with replacement of the single-center radial wave
function \mm{{\cal R}^{(j)}_{\ell}(kr)}\ with its free-electron
counterpart \mm{j_{\ell}(kr)}.

\section{Technical details\\
 of PEPD calculations}
\label{technical}

When evaluating the PEPD defined by 
Eqs.\ (\ref{psph}) and (\ref{jsphere}), we rely on the expression
\begin{eqnarray}
P^{(j)} & = &
\frac{16 \pi^2}{\sigma_{\scriptscriptstyle \text{XAS}}} \int \! 
\dstd^{2}\kkang  \: 
\frac{\dstd \sigma}{\dstd \Omega_{\bbox{k}}}  \,
\sum_{L}  \Bigl|  \sum_{L''}
\istd^{\ell''}  \beta^{(j)}_{L}(L'')  Y_{L''}^{*}(\kkang)
\Bigr|^{2}   \nonumber \\
& & \times \,
\int^{R^{(j)}_{N}}_{0} \! \! \dstd r \, r^{2} \,
\left[ R^{(j)}_{\ell}(kr) \right]^{2}  \; \; ,
\label{pocty}
\end{eqnarray}
where the \phd\ cross section is evaluated from Eq.\ (\ref{phd}) and
the amplitudes \mm{\beta^{(j)}_{L}(L'')}\ from Eq.\ (\ref{bl}).
The angular {\kkang}-integration  in (\ref{pocty}) could be performed
analytically in principle.  However, that would lead to such a
proliferation of slowly convergent sums over angular momenta, that it
is actually computationally more 
convenient to keep the {\kkang}-integral in the expression
(\ref{pocty}) and to evaluate it numerically. 

The  angular momentum sum \mm{\sum_{L}}\  in Eq.\ (\ref{pocty})
descends from  Eq.\ (\ref{rozvoj})  and converges quite quickly (it
corresponds to a multi-center expansion in the terminology of Durham
\ea).\cite{dph}  In all the cases analyzed in Sec.~\ref{aplikace}, it
was sufficient to cut this sum at \mm{\ell_{\text{max}}=3}.  On the other
hand, the sum  \mm{\sum_{L''}}, which descends into  Eq.\ (\ref{pocty})
from the expansion (\ref{bl}), corresponds to a single-center
expansion and converges only slowly.  In this work it was cut at
\mm{\ell_{\text{inc}}=30}, which we found to be a safe value.  Taking
\mm{\ell_{\text{inc}}=20}\ would still lead to an acceptable accuracy,
while increasing \mm{\ell_{\text{inc}}}\ up to 40 did not induce
significant changes with respect to the \mm{\ell_{\text{inc}}=30}\ case.

The angular integration was performed in two ``perpendicular''
spherical coordinates $\theta$, $\phi$\ as
\begin{equation}
   \int \dstd^{2}\kkang  \; \longrightarrow \;
         \int_{0}^{\pi} \dstd \theta \, \sin \theta \: 
          \int_{0}^{2 \pi} \dstd \phi \; \; .
\label{numint}
\end{equation}
In Sec.~\ref{aplikace}, fifty points were used for numerical integration
over the $\theta$\ coordinate and \mm{2\times 50\times \sin \theta}\
points for integration over $\phi$.  We checked that such a grid is
dense enough to guarantee a sufficient accuracy.

\end{document}